# Magnetocaloric effect in Gd$_3$Ir compound

Pramod Kumar[1*] and Rachana Kumar[2]
[1]Spintronics and Magnetic Materials Laboratory, Indian Institute of Information Technology Allahabad, Allahabad-211012, India
[2]National Physical Laboratory, New Deli-11012, India

Here we report magneto-thermal properties of Gd$_3$Ir compounds using magnetic and heat capacity data. We observed large isothermal entropy change($ΔS_M$), adiabatic temperature change ($ΔT_{ad}$) and relative cooling power are 4.1J/Kg, 4K and 437J/Kg on 50KOe respectively.

[*]Email Id: pkumar@iiita.ac.in

Owing to the environmental concerns and high efficiency of magnetic refrigeration as compared to the conventional vapor cycle based refrigeration technology, there has been tremendous activity in the field of magnetocaloric effect (MCE) [1-5]. The MCE is measured in terms of isothermal entropy change($\Delta S_M$) or adiabatic temperature change ($\Delta T_{ad}$) and its principles have been used for achieving sub-Kelvin temperatures [1-5]. It has been shown that the magnetic refrigeration can not only be utilized in liquid He temperature regime, but may also be used in liquid $H_2$ and room temperature regimes and, thus, there has been an increased activity for the search of potential working substances suitable for the sub-room temperature and near room temperature regimes [6, 7]. Experimental investigations have shown that materials exhibiting first order transitions possess significantly high values of $\Delta S_M$ and $\Delta T_{ad}$ and, therefore, are attractive for magnetic refrigeration applications [8-11]. However, recent investigations have also demonstrated that, apart from materials exhibiting the giant value of MCE, the compounds exhibiting moderate MCE over a wide temperature span are also important, especially for Ericsson-type of refrigeration cycle [11].

In our effort to identify potential refrigerant materials suitable for refrigeration applications we have studied the magnetic and magnetocaloric properties of intermetallic compound $Gd_3Ir$ and results are presented in this manuscript. The $Gd_3Ir$ belongs to the $R_3T$ [R= rare earth and $T$= Transition metal] family of compounds. The $R_3T$ compounds have the highest $R$ content and crystallize in the orthorhombic $Fe_3C$-type structure [12]. In $Gd_3Ir$ the Fermi level has been reported to lie in the region of positive curvature which leads to presence of strong spin fluctuation at temperature well above the ordering temperature [13]. Owing to S-shaped nature of the orbital charge density of Gd, the crystalline electric field (CEF) in $Gd_3Ir$ is expected to be negligible. Therefore, the availability of large magnetic entropy [$3Rln(2J+1)$] and negligible magnetocrystalline anisotropy [9] in $Gd_3Ir$ may lead to reasonably large MCE. Furthermore, the presence of strong spin fluctuations above the magnetic ordering temperature in $Gd_3Ir$ may drag the magnetic entropy to high temperatures and, thus, may lead to large MCE over a wide temperature span.

The polycrystalline sample was prepared by arc melting of the constituent elements of 99.9% purity in an argon atmosphere. To ensure good homogeneity, the sample was re-melted four times. The phase purity of the sample was examined by collecting the room temperature x-ray diffraction pattern using Cu-K$_\alpha$ radiation. The refinement of x-ray pattern reveals that the title compound is single phase crystallizing in monoclinic structure with lattice parameters of a=7.207(2) Å, b=9.226(2) Å, c=6.215(5) Å and β=81.53. Magnetization and heat capacity measurements were performed using the Super Conducting Quantum Interference (SQUID) magnetometer and Physical Property Measurement System (PPMS), respectively. The magnetization collected at various temperatures near the ordering temperature and the heat capacity data collected in different field have utilized to determine the magnetocaloric properties

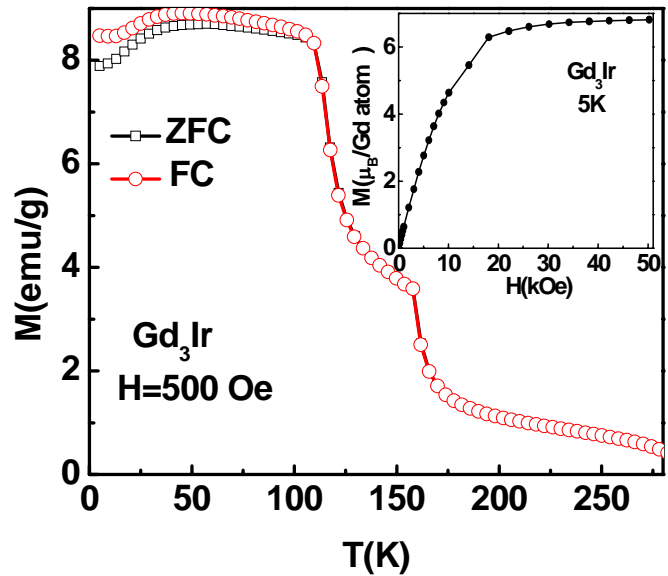

Fig.1 Temperature dependence of magnetization (M) data of $Gd_3Ir$ collected in an applied field (H) of 500 Oe under zero field cooled (ZFC) and field cooled (FC) modes. Inset shows the magnetization isotherm of $Gd_3Ir$ recorded at 5K.

Fig.1 shows the temperature dependence of magnetization (M) data collected in an applied field (H) of 500 Oe under zero fields cooled (ZFC) and field cooled (FC) conditions. We note that the M-T data of $Gd_3Ir$ show magnetic transitions at 116 K and 157 K. The magnetic transition occurring at 157 K is associated with the magnetic order-disorder transition and is denoted by $T_{ord}$ whereas the anomaly at 116 K is attributed to the spin reorientation transition and is denoted by $T_t$. A similar M-T behavior has been reported for other $R_3Ir$ compounds also [13]. Apart from presence of multiple magnetic transitions another feature worth noting in Fig. 1 is the occurrence of thermomagnetic irreversibility between the ZFC and FC magnetization data. It is well known that in ferromagnetic materials the thermomagnetic irreversibility arises from pinning of domain wall and is likely the reason for the irreversibility in present case as well [5]. It may be noted from Fig. 1 that though the order-disorder transition takes place at 157 K, the magnetization does not decrease to zero even at temperatures well above the ordering temperature. This seems to be unusual for *Gd-M* compounds and can be attributed to the strong influence of the *f* electrons of Gd ions on the *d*-electron of the transition metal which results in strong spin fluctuations through *f-d* exchange interaction [13]. Inset in Fig.1 shows the magnetic field dependence of magnetization collected at 5 K up to maximum field 50 kOe. We note that the M-H data collected for increasing and decreasing fields do not show hysteresis which implies that the compound is magnetically soft. The magnetization is found to saturate above 30 kOe and the saturation value (per Gd ion) is found to be 6.9 $\mu_B$, which is lower than the expected value of 7$\mu_B$ for $Gd^{3+}$ ion.

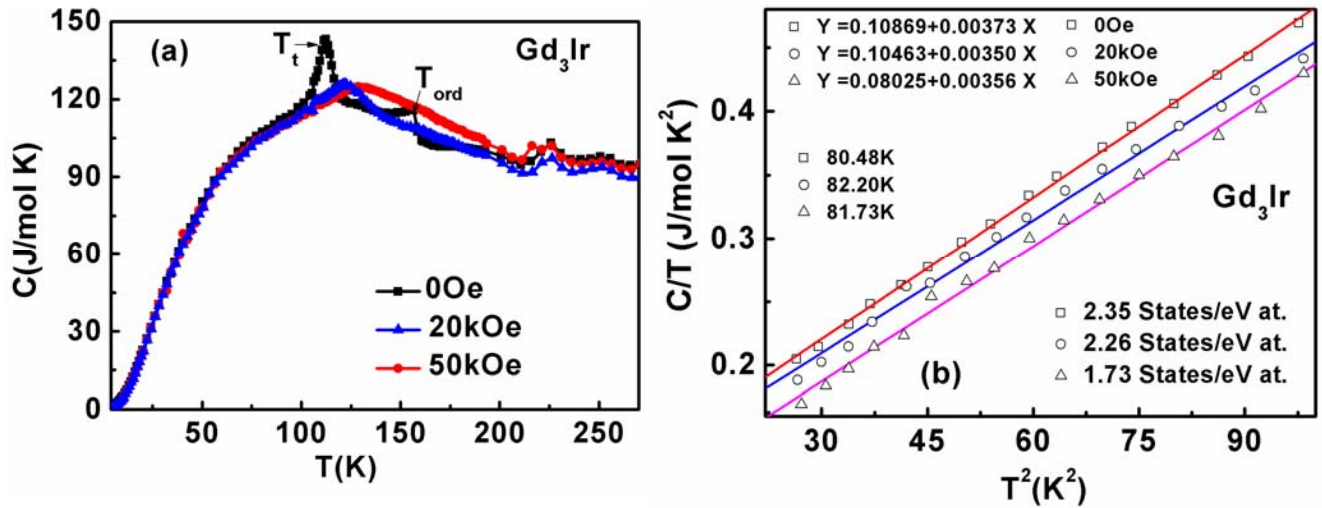

Fig.2 (a) Temperature dependence of heat capacity of $Gd_3Ir$ in applied fields of 0, 20, 50 kOe (b) The linear fit of $C/T$ vs. $T^2$ plots in different fields.

In order to further probe the nature of the magnetic state of the title compound, the heat capacity (C) measurements were performed, both under zero field and in various fields. Figure 2 (a) shows the temperature variation of heat capacity of $Gd_3Ir$ compound in 0, 20 and 50 kOe. We note that in accord with the magnetization data, the C-T data also show two anomalies at temperatures close to the spin reorientation transition temperature ($T_t$) and order-disorder transition temperature ($T_{ord}$). The anomalies associated with $T_t$ and $T_{ord}$ are found to be λ-type which implies that the magnetic transition at these temperatures is of second order character. Owing to the broad distribution of magnetic entropy the C-T anomalies become broader with increasing magnetic field.

In order to analyze the magnetic behavior of this compound, the magnetic contribution to the heat capacity ($C_M$) of $Gd_3Ir$ has been determined from the zero-field heat capacity. Following the methods reported in literature [14-16] the $C_M$ was determined from the zero-field C-T data by subtracting the lattice ($C_{lattice}$) and the electronic ($C_{ele}$) contributions. The coefficient of electronic heat capacity (γ) was determined by fitting the equation $\frac{C}{T} = \gamma + \beta T^2$ to the low temperature C-T data (see Fig. 2b). The value of γ is found to be 109 mJ/mol-$K^2$ which compares well with the value reported by Baranov et al.[17]. The γ value of nonmagnetic isostructural compound $Y_3Ir$ has been reported to be 11 mJ/mol $K^2$ [14], which is 10 times lower than the value obtained for $Gd_3Ir$. Similarly, the γ-values of $Gd_3T$ compounds with X= Co and Ni have been reported to be 110 and 100 mJ/mol-$K^2$, respectively, whereas the γ-values for their nonmagnetic analogs namely $Y_3Co$ and $Y_3Ni$ have been reported to be 15 and 14 mJ/mol-$K^2$, respectively. Thus, on the basis of these results, we infer that such a large enhancement in γ-value of $Gd_3Ir$ as compared to that of $Y_3Ir$ arises due to the presence of large spin fluctuations contribution from the d-electron subsystem of Ir.

Using the γ-value determined from the above mentioned procedure, we have also calculated the density of states using the formula $N(E_F) = \frac{3\gamma}{\kappa_B^2 \pi^2}$ [20]. The γ-values determined from the C-T data collected in applied fields of 0, 20 and 50kOe yield density of states of 2.35, 2.26, 1.73 States/eV atm., respectively. We note that for 0 and 20 kOe fields the density of states is almost equal; however, for an applied field of 50 kOe it decreases significantly. It is well known that the compounds having high density of states produce strong spin fluctuations and, therefore, for an applied field of 50 kOe, the reduction in the number of states may be taken as an indication of suppression of spin fluctuations in high applied field.

The temperature dependence of magnetic contribution to heat capacity ($C_M$) obtained after the subtraction of non magnetic contribution reveal that at $T_{ord}$ the $C_M$(T) exhibits a jump (data not shown here) of about ~70 J/mol-K. It is well known that under the mean field approximation the compounds with modulated moment exhibit a discontinuity of $10J(J+1)R/3(2J^2+2J+1)$ [where J is total angular moment and R is universal gas constant] whereas those with equal moment configuration exhibit a discontinuity of $5J(J+1)R/(2J^2+2J+1)$ [21]. Thus, in the present case we expect a discontinuity of ~40 J/mol-K for and amplitude modulated configuration whereas for equal moment configuration a jump of ~60 J/mol-K is expected. The fact that Gd$_3$Ir exhibits a discontinuity of ~70 J/mol-K thus indicates that the title compound assumes an equal moment [21]. The difference between the observed (~70 J/mol-K) and maximum expected (~60 J/mol-K) discontinuities may be attributed to the spin fluctuation. We note that in case of Gd$_3$Ir the $C_M$ is quite significant even up to room temperature, which is much higher temperature compared to the order-disorder transition temperature ($T_{ord}=$ 157 K). This is quite unusual especially because crystal field effect is expected to be absent in this Gd-based compound and may be attributed to presence of strong spin fluctuations at temperatures well above $T_{ord}$. It may be recalled here that arising from spin fluctuations the *M-T* data also showed the presence of considerable magnetization at temperatures well above $T_{ord}$. In Gd$_3$Ir, based on the magnetization studies Talik and Slebarski have also proposed the presence of strong spin fluctuations at temperatures well above ordering temperatures [13].

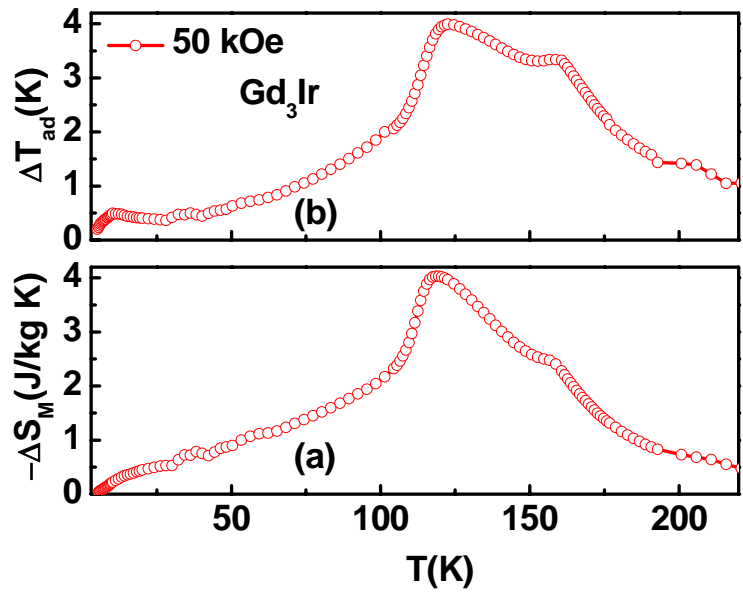

Fig.3 Temperature dependence of (a) Isothermal magnetic entropy change (-$\Delta S_M$) and (b) adiabatic temperature change ($\Delta T_{ad}$) for a field change of 50 kOe.

The magnetocaloric properties of this compound have been determined in terms of isothermal magnetic entropy change ($\Delta S_M$) and adiabatic temperature change ($\Delta T_{ad}$) using the heat capacity data. The former has been calculated from both M-H-T and C-H-T data whereas the later has been calculated using the *C-H-T* data [24,25]. The temperature dependence of the MCE thus obtained is shown in Figure 3. It can be seen from the figure 3a that the $\tau S_M$ *vs. T* plot of the compound shows a maximum near the $T_{ord}$ and the temperature variation of $\Delta T_{ad}$ follows the variations seen in the entropy change (see Fig. 3b). It may be noted that owing to presence of two magnetic transitions (see Fig. 1) at $T_t$ and $T_{ord}$, temperature variation MCE of $Gd_3Ir$ shows double-peak structure. For a field change (ΔH) of 50 kOe the $\Delta S_M$ values corresponding to magnetic transitions at $T_t$ and $T_{ord}$ are found to be 4.1 and 2.5 J/kg-K, respectively whereas the values of adiabatic temperature change at these temperatures are found to be 4.1 and 3.5 K, respectively. It may be mentioned here that $\Delta T_{ad}^{max}$ value, for ΔH= 50 kOe in $RNi_2$ compounds varies in the range of 3.5 - 9 K [31] whereas for $(Er,Dy)Al_2$ compounds for the same field change, this varies between 7 and 11 K [32]. The $\Delta T_{ad}^{max}$ of GdPd, which is an active magnetic regenerator material used in the low temperature rage of magnetic refrigerators for liquefaction of hydrogen is 8.7 K, for a field change of 70 kOe [2, 33]. Therefore, a comparison of the $\Delta T_{ad}^{max}$ values of $Gd_3Ir$ with that of the potential refrigerant materials indicates that this compound may find application as magnetic refrigerant at low temperatures. Another criterion which is used to grade the magnetic refrigerant materials is relative cooling power (RCP). The RCP is measure of heat transfer between hot and cold sinks in an ideal refrigeration cycle and is defined as RCP=-$\Delta S_M^{max} \times \delta T_{FWHM}$ (where $\delta T_{FWHM}$ is the full width at half maximum of $\tau S_M$ *vs. T* plot). For ΔH= 50 kOe, the RCP of Gd3Ir is found to be 437J/kg. Thus we note that the

RCP of the present compound (like LaFe$_{11.4}$Si$_{1.6}$ is 420 J/Kg) is comparable to many of the potential magnetic refrigerants which makes this compound potential candidate for refrigeration applications around 150 K.

In conclusion, we have studied the magneto-thermal properties of Gd$_3$Ir compound. The magnetization data reveals that the Ir-sublattice is nearly nonmagnetic. The magnetocaloric property of this compound is found to be comparable to that of many potential refrigerant materials. The spin fluctuations lead to considerable MCE at temperatures well above ordering temperatures, which leads to the occurrence of large RCP value in the present case.